\documentclass[%
 aip,
 apl,
 amsmath,amssymb,
preprint,%
]{revtex4-1}

\usepackage{graphicx}
\usepackage{dcolumn}
\usepackage{bm}
\usepackage[mathlines]{lineno}
\usepackage{soul,color}
\usepackage[utf8]{inputenc}
\usepackage[T1]{fontenc}
\usepackage{mathptmx}
\usepackage{etoolbox}
\usepackage{float}

\begin{document}

\title{Significant ballistic thermal transport across graphene layers: effect of nanoholes and lithium intercalation}
\author{John Crosby} \email{johncrosby@unr.edu}
\author{Haoran Cui} \email{hcui@unr.edu; co-first author with equal contribution; co-corresponding author}
\author{Yan Wang}
 \email{yanwang@unr.edu; co-corresonding author}
\affiliation{Department of Mechanical Engineering, University of Nevada, Reno, Reno, NV 89557, USA}%
\date{\today}
\begin{abstract}
Porous graphene and graphite are increasingly utilized in electrochemical energy storage and solar-thermal applications due to their unique structural and thermal properties. In this study, we conduct a comprehensive analysis of the lattice thermal transport and spectral phonon characteristics of holey graphite and multilayer graphene. Our results reveal that phonon modes propagating obliquely with respect to the graphene basal planes are the primary contributors to cross-plane thermal transport. These modes exhibit a predominantly ballistic nature, resulting in an almost linear increase in cross-plane thermal conductivity with the number of layers. The presence of nanoholes in graphene induces a broadband suppression of cross-plane phonon transport, whereas lithium ion intercalation shows potential to enhance it. These findings provide critical insights into the mechanisms governing cross-plane heat conduction in key graphene-based structures, offering valuable guidance for thermal management and engineering of van der Waals materials.
\end{abstract}

\maketitle

Van der Waals (vdW) materials comprise stacked atomic layers held together by weak van der Waals forces, while strong covalent or ionic bonds dominate within each layer. These materials exhibit remarkable versatility and tunability in their electronic, optical, thermal, and mechanical properties, making them highly attractive for a wide range of applications. Their physical behavior is particularly sensitive to thermodynamic conditions, chemical composition, structural defects, stacking order, and layer number---features that have led to significant advancements in nanoelectronics, photonics, and quantum technologies. Among vdW materials, graphene, a single layer of sp$^2$-hybridized carbon atoms arranged in a honeycomb lattice, has emerged as a model system. A key attribute of graphene's tunability lies in the strong dependence of its electronic and optical properties on stacking sequence~\cite{mak2010electronic,bao2011stacking,Cao2018Nature} and layer number~\cite{mak2010evolution}.

In recent years, multilayer graphene has been engineered in various ways to meet the demands of specific applications. For instance, nano-holey graphene architectures---graphene layers perforated with nanoscale holes---have been developed to enhance lithium-ion transport across layers~\cite{wang2024manufacturing}. These structures exhibit exceptional ion conductivity and energy storage capabilities, positioning them as promising candidates for use in supercapacitors and lithium-ion batteries. Moreover, graphite and multilayer graphene have long served as electrode materials in lithium-ion batteries, where lithium ions intercalate between graphene layers~\cite{Qian2016ACS}. This intercalation process inherently modifies the thermal conductivity ($\kappa$) of otherwise pristine multilayer graphene.

The thermal transport properties of pristine graphite (PG) have been extensively explored through both experimental~\cite{ho1974thermal,fugallo2014thermal} and theoretical methods~\cite{lindsay2011flexural,nika2012two,wang2014two,qian2016anisotropic}. The in-plane thermal conductivity ($\kappa_{\parallel}$) of single- and few-layer graphene, as well as graphite, has been characterized using various measurement techniques~\cite{pop2012thermal,nika2012two} and computational approaches including first-principles calculations and atomistic simulations~\cite{pop2012thermal,wang2014two}. A consistent finding is that $\kappa_{\parallel}$ decreases as the number of layers increases, eventually saturating at the graphite limit~\cite{ghosh2010dimensional,lindsay2011flexural}. This reduction is primarily attributed to the breakdown of mirror symmetry in multilayer structures, which enables additional anharmonic scattering mechanisms that are otherwise prohibited in single-layer graphene ~\cite{lindsay2011flexural}.

In contrast, the cross-plane thermal conductivity ($\kappa_{\perp}$) remains more difficult to measure and is less well understood. Consequently, the dependence of cross-plane phonon transport on layer number is not yet fully resolved~\cite{wei2011interfacial,ni2013significant,shen2025point}. Atomistic simulations have reported a notable increase in $\kappa_{\perp}$ with increasing graphene layers in pristine multilayer structures~\cite{wei2011interfacial}, with no clear saturation observed. This trend implies that phonons contributing to cross-plane transport may possess mean free paths extending beyond the micrometer scale. Despite the intriguing nature of this phenomenon, further investigation remains limited. Prior studies have shown that lithium-ion intercalation and nanopore introduction can significantly influence in-plane thermal transport~\cite{qian2016anisotropic,cui2024elucidating,wang2014two}, yet their effects on cross-plane conductivity have not been systematically quantified.

Despite these advances, several critical questions regarding cross-plane phonon transport in multilayer graphene remain unanswered. Specifically: How do nanoscale holes and their spatial arrangement affect $\kappa_{\perp}$? How does lithium-ion intercalation alter the dependence of $\kappa_{\perp}$ on layer number? Which phonon modes are most sensitive to stacking variations, and how do they contribute to the observed changes in thermal transport? In this work, we address these questions using molecular dynamics (MD) simulations in conjunction with spectral phonon analysis.

\begin{figure}
    \centering
    \includegraphics[width=0.5\linewidth, keepaspectratio]{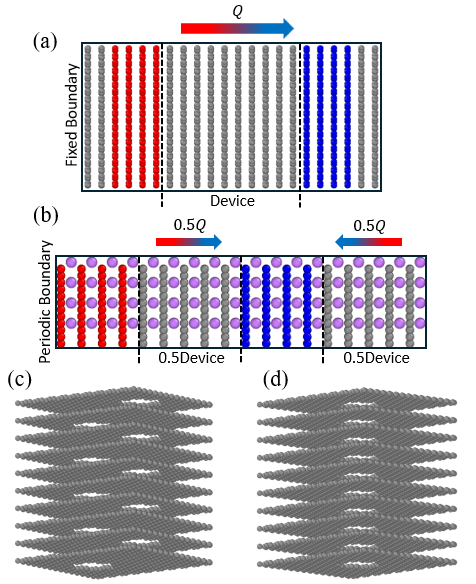}
    \caption{(a) and (b): NEMD simulation setups for calculating cross-plane thermal conductivity ($\kappa_{\perp}$) of PG, AHG, and SHG (a) and LiC$_6$ (b). Setup (b) employs fully periodic boundary conditions for LiC$_6$ to avoid artificial truncation of long-range Coulombic interactions introduced by fixed boundaries in (a). (c) and (d): Atomic structures of AHG and SHG, respectively. The AHG structure features through-holes, while the SHG structure contains staggered holes that do not form continuous perforations.}
    \label{fig:figure1}
\end{figure}

To investigate the dependence of $\kappa_{\perp}$ on the number of graphene layers, we perform non-equilibrium molecular dynamics (NEMD) simulations using the LAMMPS package~\cite{LAMMPS}. Four representative graphite-based structures are examined: pristine graphite (PG), holey graphite with aligned holes (AHG), holey graphite with staggered holes (SHG), and lithium-ion intercalated graphite (LiC$_6$), as illustrated in Fig.~\ref{fig:figure1}. This comparative analysis aims to elucidate the phonon transport mechanisms governing $\kappa_{\perp}$ in vdW bonded systems and to inform the thermal design of layered carbon-based materials.

For PG, AHG, and SHG, the cross-sectional area of each graphene layer is fixed at \(A_g = 42.6~\text{\AA} \times 34.4~\text{\AA}\). The AHG and SHG configurations are generated by removing 7\% of atoms per layer to create circular nanopores. While previous studies have shown that smaller simulation domains are sufficient for convergence in both EMD and NEMD simulations, a larger cross-sectional area is employed here for PG and holey graphite to prevent unrealistic interlayer slipping due to weak vdW forces. For LiC$_6$, a smaller cross-sectional area of \(A_{\text{LiC}_6} = 15.9~\text{\AA} \times 15.9~\text{\AA}\) is used, which is sufficient to maintain structural integrity and achieve convergence. Each interlayer space in LiC$_6$ is uniformly intercalated with 16 lithium ions, yielding a stoichiometry consistent with Li:C = 1:6, in agreement with prior work~\cite{Qian2016ACS}.

Intralayer carbon-carbon interactions are modeled using the Tersoff potential~\cite{Kinaci2012PRB}, while interlayer vdW interactions are described by the Lennard-Jones (LJ) potential:
\begin{equation}
    E = 4\epsilon \left[\left(\frac{\sigma}{r}\right)^{12} - \left(\frac{\sigma}{r}\right)^6\right],
\end{equation}
where $r$ is the interatomic distance and the cutoff radius $r_c$ is set to 10~\text{\AA}. In the LiC$_6$ system, the LJ potential is combined with a Coulombic term to model long-range electrostatic interactions between carbon and lithium atoms:
\begin{equation}
    E = 4\epsilon \left[\left(\frac{\sigma}{r}\right)^{12} - \left(\frac{\sigma}{r}\right)^6\right] + \frac{C q_i q_j}{\epsilon r},
\end{equation}
where $q_i$ and $q_j$ are the charges of the interacting atoms ($q_C = -0.147e$, $q_{Li} = 0.882e$), and $C$ is the Coulomb constant.

The simulation setup for PG, AHG, and SHG is shown in Fig.~\ref{fig:figure1}(a), which includes two fixed layers at each end to mimic rigid boundaries and 50-layer hot and cold reservoirs to capture both short- and long-wavelength phonons (only four layers are shown in the figure for clarity). The system is equilibrated using a Nosé-Hoover thermostat and barostat~\cite{Nose1984JCP,Hoover1985PRA} in two $NPT$ stages: (1) a linear ramp from 5~K to 300~K over 0.5~ns and (2) equilibration at 300~K for an additional 0.5~ns. All simulations employ a time step of 0.5~fs.

Following equilibration, the NEMD simulation proceeds with plain time integration. A temperature gradient is imposed by velocity rescaling within the hot and cold reservoirs at $T_{hot} = 320$~K and $T_{cold} = 280$~K, respectively, generating a thermal bias of $\Delta T = 40$~K. Heat flow is induced across the structure for 20~ns, during which the energy added to the hot bath (${Q}_{in}$) and removed from the cold bath (${Q}_{out}$) are recorded. The heat current is calculated from the linear fit of ${Q}_{in}$ and ${Q}_{out}$ with respect to time. The cross-plane thermal conductivity is then obtained as
\begin{equation}
    \kappa_{\perp} = \frac{\dot{Q}L}{A\Delta T},
    \label{eq:fourier}
\end{equation}
where $\dot{Q} = \frac{1}{2} \left( \dot{Q}_{in} + \dot{Q}_{out} \right)$ is the mean steady-state heat current and $L$ is the distance between the hot bath and cold bath. 

A different approach is necessary for LiC$_6$ due to the influence of long-range Coulomb interactions, which can lead to spurious energy transfer across periodic boundaries. To address this, a fully periodic configuration (Fig.~\ref{fig:figure1}b) is used, where the heat is symmetrically conducted along both the $+z$ and $-z$ directions. This eliminates fixed layers and allows for computational efficiency while ensuring physical accuracy. $\kappa_{\perp}$ is computed using Eq.~\ref{eq:fourier} but with a factor of 0.5 to account for the symmetric splitting of $\dot{Q}$ in two directions. 

Equilibrium molecular dynamics (EMD) simulations are also conducted to determine the bulk-limit $\kappa_{\perp}$ using the Green-Kubo formalism. Twelve-layer configurations of PG, AHG, SHG, and LiC$_6$ are simulated. Following a structural relaxation process similar to the NEMD procedure, the instantaneous heat flux is sampled every 10 time steps (5~fs) in the microcanonical ensemble over a total duration of 30~ns. In Supplementary Materials, we show the convergence of Green-Kubo curves with respect to the number of graphene layers, confirming that 12 layers are sufficient for obtaining bulk-limit $\kappa_{\perp}$.

\begin{figure}
    \centering
    \includegraphics[width=0.8\textwidth, keepaspectratio]{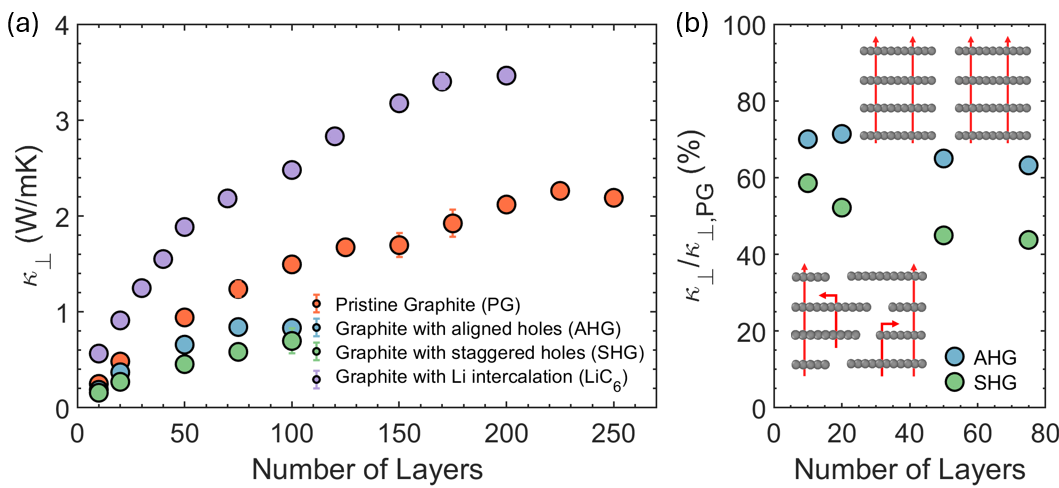}
    \caption{ (a) Cross-plane thermal conductivity $\kappa_{\perp}$ of PG, AHG, SHG, and LiC$_6$ as a function of layer number. Bulk $\kappa_{\perp}$ values obtained via EMD are $6.9 \pm 0.4$ W/m-K, $3.0 \pm 0.1$ W/m-K, $2.3 \pm 0.2$ W/m-K, and $7.13 \pm 0.16$ W/m-K for PG, AHG, SHG, and LiC$_6$, respectively. (b) Percent ratio of $\kappa_{\perp}$ in AHG and SHG to $\kappa_{\perp}$ in PG as a function of layer number. The insets illustrate the representative cross-plane phonon propagation paths in AHG and SHG, showing the phonons are more often blocked by the staggered nanoholes in SHG.}
    \label{fig:k}
\end{figure}

Figure~\ref{fig:k}a shows the cross-plane thermal conductivity $\kappa_{\perp}$ as a function of the number of layers for PG, AHG, SHG, and lithium-intercalated graphite LiC$_{6}$. In all cases, $\kappa_{\perp}$ increases significantly with the number of layers. For PG, in particular, $\kappa_{\perp}$ increases nearly linearly with layer number up to 80 layers, indicating a strong ballistic transport regime.

Furthermore, as shown in Fig.~\ref{fig:k}a, the presence of nanoholes significantly suppresses cross-plane thermal transport. Across all simulated thicknesses, the $\kappa_{\perp}$ of holey graphite structures remains consistently lower than that of pristine graphite. EMD simulations corroborate the NEMD results and enable estimation of bulk-limit behavior. The bulk-limit $\kappa_{\perp}$ of PG is $6.9 \pm 0.4$~W/m-K, agreeing with previously reported values~\cite{Ho1972JPCRefData,Schmidt2008RSI}, while AHG and SHG exhibit significantly lower bulk-limit thermal conductivities of $3.0 \pm 0.1$~W/m-K and $2.3 \pm 0.2$~W/m-K, respectively, confirming the detrimental impact of nanoholes on interlayer heat conduction.

The $\kappa_{\perp}$ of SHG is substantially lower than that of AHG. This is attributed to the staggered configuration of holes in SHG, which increases the likelihood that a cross-plane or obliquely propagating phonon encounters a void and is scattered, as schematically compared in the inset of Fig.~\ref{fig:k}b. To further highlight the suppression effect, Fig.~\ref{fig:k}b presents the ratio of $\kappa_{\perp}$ in holey graphite to that in PG with the same number of layers. The $\kappa_{\perp}$ of AHG is approximately 60–70\% that of PG, while SHG retains only 40–60\% of PG’s $\kappa_{\perp}$. The ratios decline further with increasing layer count, approaching their respective bulk limits. As a result, both AHG and SHG display a much weaker layer-number dependence in $\kappa_{\perp}$ compared to PG, consistent with the trends in Fig.~\ref{fig:k}a.

In contrast to holey structures, lithium intercalation significantly enhances cross-plane thermal conductivity. EMD simulations show that the bulk $\kappa_{\perp}$ of LiC$_6$ reaches $7.13 \pm 0.16$~W/m-K, surpassing that of PG. This observation supports earlier studies reporting increased $\kappa_{\perp}$ at a 1:6 Li:C ratio~\cite{Wenjing2020AIP,Zhiyong2018JPC}, although suppression effects under similar conditions have also been reported~\cite{Qian2016ACS}. It is worth noting that our simulated LiC$_6$ structure assumes an ideal periodic lattice with uniformly distributed Li ions. In realistic systems, disorder in Li ion arrangement and deviations from the 1:6 stoichiometry may reduce the observed thermal conductivity enhancement.

To elucidate the mechanism of cross-plane thermal transport and the effect of nanoholes, we compute the spectral thermal conductivity, $\kappa(\omega)$, i.e., the decomposition of $\kappa_{\perp}$ into phonon frequency contributions. Specifically, the spectral phonon heat flux $Q(\omega)$ is extracted from NEMD simulations using the following expression~\cite{saaskilahti2014PRB,saaskilahti2015PRB}:
\begin{equation}
    Q(\omega)=\sum_{i\in\tilde{L}}\sum_{j\in\tilde{R}}	\left( -\frac{2}{t_{\mathrm{simu}}\omega}\sum_{\alpha,\beta}\mathrm{Im}\left\langle\hat{v}_i^\alpha(\omega)^*K_{ij}^{\alpha\beta}\hat{v}_j^\beta(\omega)\right\rangle \right), 
    \label{eqn:spectral_q}
\end{equation}
where $i$ and $j$ are atomic indices in the left ($\tilde{L}$) and right ($\tilde{R}$) halves of a central cross-section, $t_{\mathrm{simu}}$ is the total simulation time, $\alpha$ and $\beta$ denote Cartesian directions ($x$, $y$, $z$), $\hat{v}$ is the Fourier-transformed atomic velocity, $^*$ denotes complex conjugation, and $K_{ij}^{\alpha\beta}$ is the force constant matrix.

Heat flux contributions can be further decomposed by vibrational polarization~\cite{CuiPRB}:
\begin{align}
    Q_{x}(\omega) &=\sum_{i\in\tilde{L}}\sum_{j\in\tilde{R}} \left( -\frac{2}{t_{\mathrm{simu}}\omega} \sum_{\beta=x,y,z} \mathrm{Im} \left\langle \hat{v}_i^x(\omega)^* K_{ij}^{x\beta} \hat{v}_j^\beta(\omega) \right\rangle \right), \label{eqn:spectral_qx} \\
    Q_{y}(\omega) &=\sum_{i\in\tilde{L}}\sum_{j\in\tilde{R}} \left( -\frac{2}{t_{\mathrm{simu}}\omega} \sum_{\beta=x,y,z} \mathrm{Im} \left\langle \hat{v}_i^y(\omega)^* K_{ij}^{y\beta} \hat{v}_j^\beta(\omega) \right\rangle \right), \label{eqn:spectral_qy} \\
    Q_{z}(\omega) &=\sum_{i\in\tilde{L}}\sum_{j\in\tilde{R}} \left( -\frac{2}{t_{\mathrm{simu}}\omega} \sum_{\beta=x,y,z} \mathrm{Im} \left\langle \hat{v}_i^z(\omega)^* K_{ij}^{z\beta} \hat{v}_j^\beta(\omega) \right\rangle \right). \label{eqn:spectral_qz}
\end{align}

To compare structures of varying thicknesses, we normalize $Q(\omega)$ to obtain spectral thermal conductivity:
\begin{equation}
    \kappa(\omega)= \frac{\kappa}{\int_{0}^{\infty}\,Q(\omega)d\omega}\cdot Q(\omega).
    \label{eqn:spectral_k}
\end{equation}

\begin{figure}
    \centering
    \includegraphics[width=\textwidth, keepaspectratio]{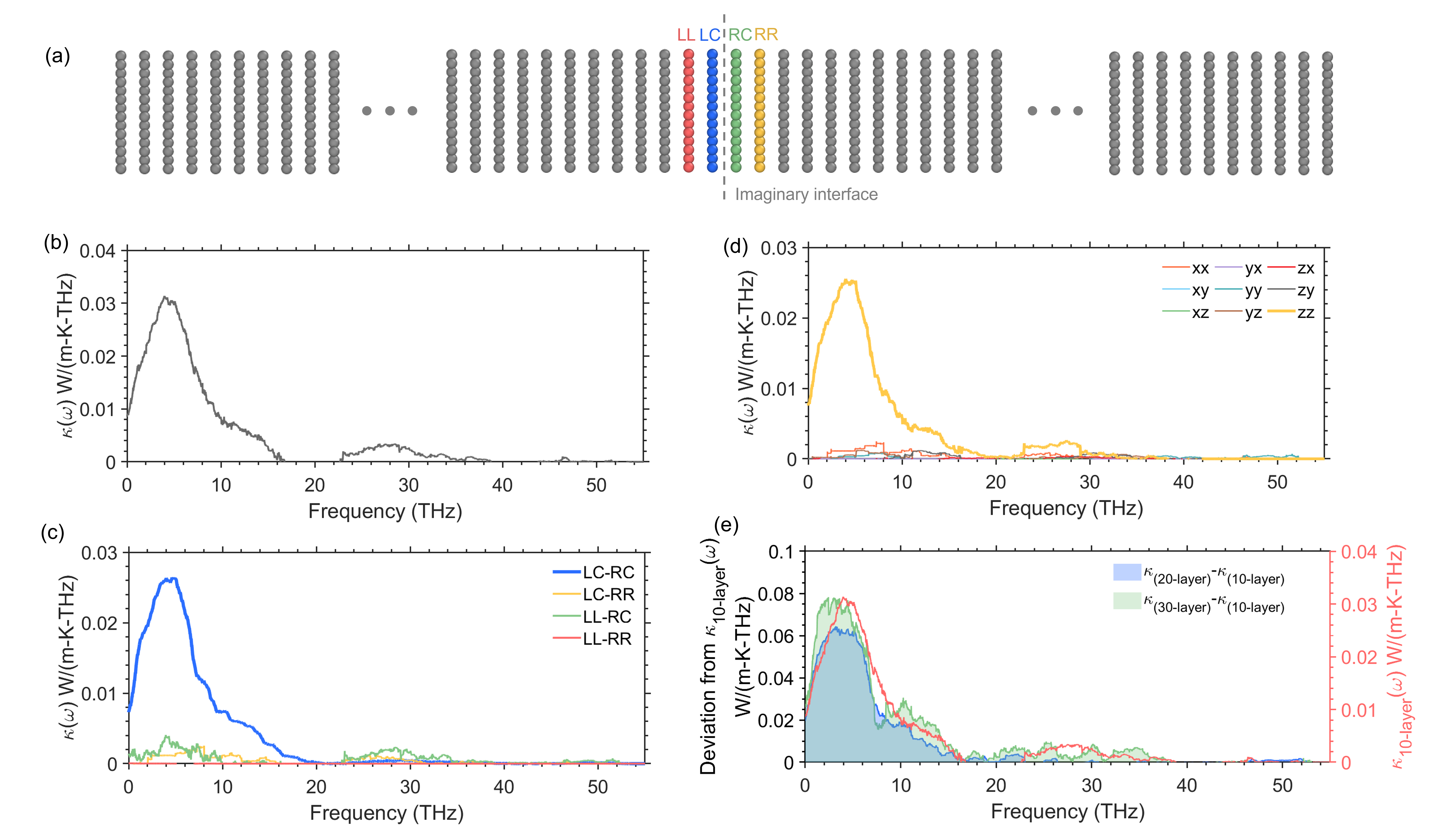}
    \caption{Spectral thermal conductivity $\kappa(\omega)$ for cross-plane thermal transport in pristine graphite (PG). (a) Schematic showing four neighboring graphene layers used in NEMD simulations to compute $\kappa(\omega)$. The dashed vertical line marks the imaginary interface for evaluating spectral heat flux $Q(\omega)$ via Eq.~\ref{eqn:spectral_q}. (b) $\kappa(\omega)$ of 10-layer PG. (c) Decomposition of $\kappa(\omega)$ by interlayer separation: LC–RC (nearest neighbors, one interlayer spacing), LC–RR and LL–RC (second-nearest neighbors, two spacings), and LL–RR (third-nearest neighbors, three spacings). (d) Decomposition of $\kappa(\omega)$ by vibrational polarization. Red curve (right axis): $\kappa(\omega)$ for 10-layer PG; shaded regions (left axis) show deviations in the 20-layer and 30-layer PG spectra relative to the 10-layer case.}
    \label{fig:spectral_pristine}
\end{figure}

Figure~\ref{fig:spectral_pristine}a shows the schematic setup for $\kappa(\omega)$ calculations. Four graphene layers located near the center of the device are selected: LL (left-left), LC (left-center), RC (right-center), and RR (right-right). Thermal conductivity is computed for layer pairs with varying separations: LC--RC (adjacent), LL--RC and LC--RR (two-layer separation), and LL--RR (three-layer separation, $\sim$10.05~\AA). Given the 10~\AA~cutoff for the Lennard-Jones potential, minimal heat exchange is expected beyond this range.

Figure~\ref{fig:spectral_pristine}b reveals that phonons with frequencies below 17~THz dominate cross-plane heat transport in pristine graphite, while higher-frequency modes (22--38~THz) contribute only modestly. As illustrated in Fig.~\ref{fig:spectral_pristine}c, the majority of heat is carried by phonons interacting between adjacent layers (LC--RC), whereas contributions from second- and third-nearest neighbors are negligible. This observation has important implications for understanding the impact of nanoholes on cross-plane phonon transport: the presence of a hole in a single graphene layer can severely impede thermal transport, as phonons are unable to effectively transmit across more than one interlayer spacing.

Figure~\ref{fig:spectral_pristine}d presents the polarization decomposition of $\kappa(\omega)$. The dominant contribution arises from out-of-plane ($zz$) vibrations, while mixed ($zx$, $zy$) and purely in-plane ($xx$, $yy$) contributions are minimal. This indicates that in-plane longitudinal and transverse modes do not significantly contribute to $\kappa_{\perp}$. However, it is not solely the flexural modes within each graphene layer that are responsible—later sections reveal that obliquely propagating acoustic phonons are the primary heat carriers.

Figure~\ref{fig:spectral_pristine}e illustrates the evolution of $\kappa(\omega)$ with increasing layer number. The nearly uniform spectral enhancement observed from 10-layer to 20-layer and 30-layer PG indicates ballistic phonon transport across the dominant frequency range. This consistent increase in spectral thermal conductivity explains the almost linear growth of the overall cross-plane thermal conductivity $\kappa_{\perp}$ with the number of graphene layers.

\begin{figure}
    \centering
    \includegraphics[width=0.75\textwidth, keepaspectratio]{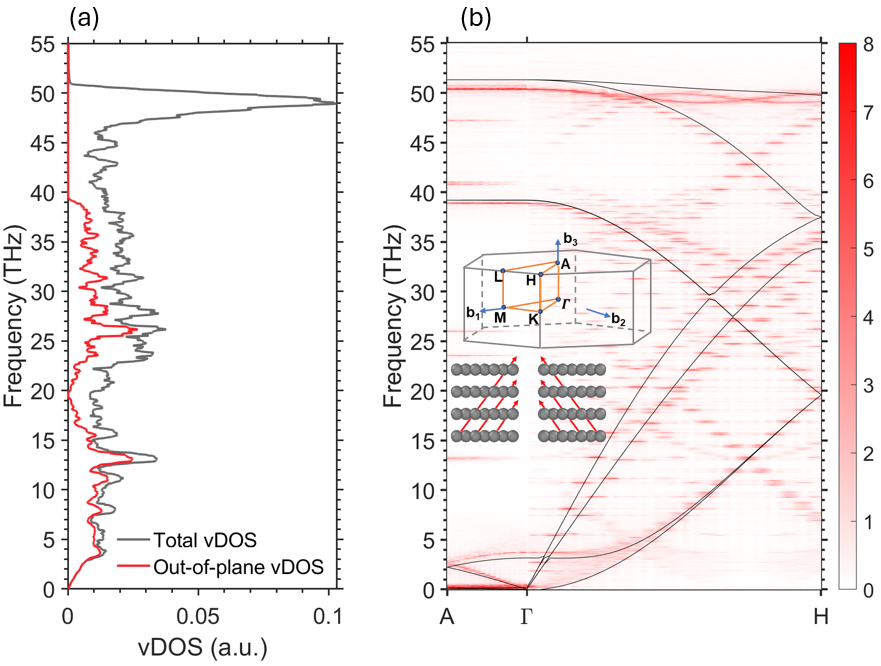}
    \caption{Phonon properties of 10-layer pristine graphite. (a) Vibrational density of states. (b) Phonon dispersion relations obtained from spectral energy density (heat map) and lattice dynamics (black lines) along the $A$--$\Gamma$--$H$ path. The upper inset shows the first Brillouin zone of graphite. The lower inset schematically illustrates transport paths of oblique phonons, highlighting their susceptibility to scattering by through-holes.}
    \label{fig:spectral_vDOS_SED}
\end{figure}

To further elucidate the primary contributors to cross-plane thermal transport, we analyze the detailed phonon spectra, including the vibrational density of states (vDOS) and phonon dispersion relations. As shown in Figure~\ref{fig:spectral_vDOS_SED}a, the vDOS is dominated by out-of-plane modes within the 0--17~THz range. Figure~\ref{fig:spectral_vDOS_SED}b compares the phonon dispersion relations obtained from lattice dynamics calculations and those extracted via spectral energy density (SED) analysis~\cite{SEDPRB}. Along the $\Gamma$--$A$ direction (normal to the basal plane), the acoustic modes are confined to low frequencies below 3~THz due to the weak interlayer shear coupling characteristic of this vdW-bonded material. Consequently, phonons propagating in the cross-plane (normal) direction contribute only to the 0--3~THz portion of the spectral thermal conductivity $\kappa(\omega)$ in Fig.~\ref{fig:spectral_pristine}b.

To identify the phonons responsible for the dominant 3--17~THz contribution to $\kappa(\omega)$, we examine dispersions along oblique directions. As illustrated in Fig.~\ref{fig:spectral_vDOS_SED}b, the acoustic branches along the oblique $\Gamma$--$H$ direction extend to much higher frequencies, corresponding well with the main spectral range of $\kappa(\omega)$. This indicates that cross-plane heat transport is primarily mediated by obliquely propagating acoustic phonons.

Further evidence supporting the dominant role of oblique phonons is found in Fig.~\ref{fig:k}b. The AHG structures contain through-holes occupying approximately 7\% of the graphene layer's cross-sectional area. If cross-plane transport were governed mainly by phonons propagating strictly in the normal direction, one would expect a corresponding $\sim$7\% reduction in $\kappa_{\perp}$. However, Fig.~\ref{fig:k}b shows that $\kappa_{\perp}$ in AHG structures is reduced by 30--40\% compared to their pristine graphite counterparts. As schematically illustrated in Fig.~\ref{fig:spectral_vDOS_SED}b, this pronounced reduction arises because obliquely propagating phonons have a significantly higher probability of encountering and being scattered by the through-holes, thereby diminishing $\kappa_{\perp}$ more substantially.

\begin{figure}
    \centering
    \includegraphics[width=0.6\textwidth, keepaspectratio]{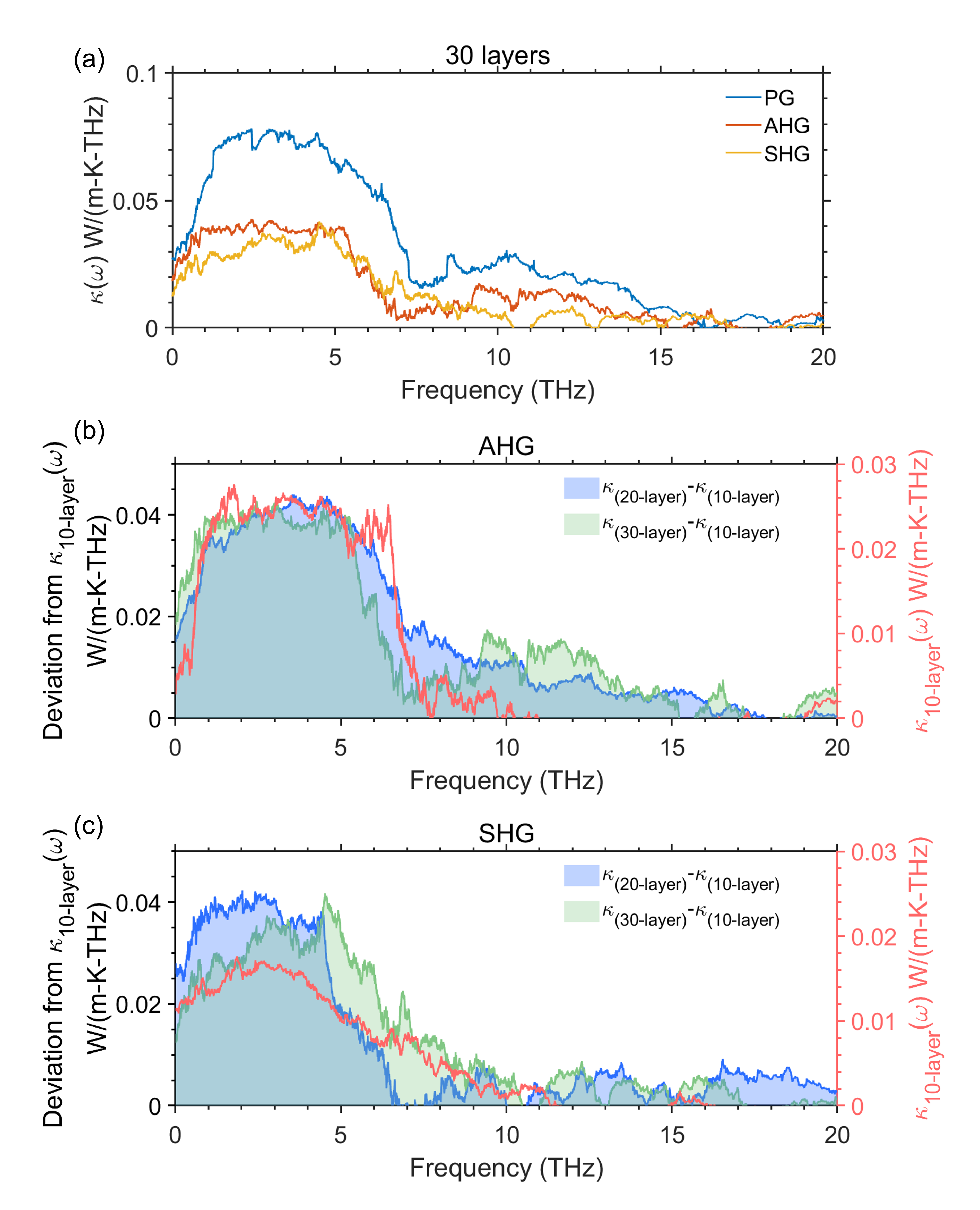}
    \caption{Spectral thermal conductivity $\kappa(\omega)$ for cross-plane thermal transport in pristine graphite (PG), aligned-hole graphite (AHG), and staggered-hole graphite (SHG). (a) Comparison of $\kappa(\omega)$ for 30-layer samples of PG, AHG, and SHG. (b) Red curve (right axis): $\kappa(\omega)$ for cross-plane thermal transport in 10-layer AHG; shaded regions, which are plotted against the left axis, indicate the deviation of the 20-layer and 30-layer AHG spectra from that of the 10-layer case. (c) Same as panel (b), but for SHG.
}
    \label{fig:spectral_all}
\end{figure}

To understand the effect of nanoholes on thermal transport contributed by each phonon modes, we also perform spectral thermal conductivity analysis of AGH and SHG structures. Figure~\ref{fig:spectral_all}a compares $\kappa(\omega)$ for 30-layer PG, AHG, and SHG. Evidently, nanoholes reduce $\kappa(\omega)$ across the entire spectrum, with SHG exhibiting the greatest suppression due to its staggered hole configuration. Unlike AHG, SHG obstructs direct cross-plane pathways, increasing phonon scattering. Figures~\ref{fig:spectral_all}b and~\ref{fig:spectral_all}c show that $\kappa(\omega)$ increases with layer number in both AHG and SHG, consistent with a ballistic transport regime despite structural disorder.

To summarize, in this work, we systematically investigated the cross-plane thermal transport properties of pristine graphite, holey graphite, and lithium-intercalated graphite (LiC$_6$) using molecular dynamics simulations. Our results demonstrate a strong dependence of $\kappa_{\perp}$ on the number of layers for all graphite-based systems, highlighting significant ballistic contributions to thermal transport even in multilayer structures extending up to 300 layers. In these graphite structures, $\kappa_{\perp}$ exhibits an almost linear increase with the number of layers, indicating long phonon mean free paths and minimal scattering across the graphene basal planes. Spectral thermal conductivity analyses further revealed that this behavior originates from phonon modes below 17~THz, with dominant contributions from obliquely propagating acoustic phonons. These phonons exhibit significant group velocities and extended mean-free-paths, resulting in efficient cross-plane thermal transport. The introduction of nanoholes in AHG and SHG significantly suppresses $\kappa_{\perp}$ by directly disrupting cross-plane phonon transport. The suppression is more pronounced in SHG due to the disruption of cross-plane continuity. Interestingly, lithium intercalation at a 1:6 Li-to-C ratio leads to an enhancement in $\kappa_{\perp}$, which is attributed to the increased interlayer coupling induced by lithium ions. Overall, this study provides critical insights into the phonon transport mechanisms in layered graphite systems and offers guidance for engineering the $\kappa_{\perp}$ of graphite through structural modifications such as nanohole patterning and intercalation. 

\section*{Acknowledgments}
The authors acknowledge the financial support from the National Science Foundation (CBET-2211696). John Crosby thanks the support of the National Aeronautics and Space Administration (NASA) for the Nevada Space Grant Undergraduate Scholarship through the Nevada Space Grant Consortium. The authors acknowledge the support of Research and Innovation and the Cyberinfrastructure Team in the Office of Information Technology at the University of Nevada, Reno, for facilitation and access to the Pronghorn High-Performance Computing Cluster.

\clearpage

\bibliography{aipsamp}
\end{document}